\documentclass[aps,pra,twocolumn,groupedaddress,showpacs]{revtex4}
\usepackage{amssymb}
\usepackage{amsmath}
\usepackage{graphicx}

\begin{document}

\title{The non-conditional nature of the Cerf-Adami inequalities and implications for thermodynamics}

\author{Ian T. Durham}
\email[]{idurham@anselm.edu}
\affiliation{Department of Physics, Saint Anselm College, Manchester, NH 03102}
\date{\today}

\begin{abstract}
We show that the Cerf-Adami inequalities do not necessarily depend on conditional entropies nor any reference to Markov chains.  While the latter are not explicit in the original form, they are often implied in certain derivations.  We also show that these inequalities are intimately related to at least one interpretation of the second law of thermodynamics.  The combination of these results provides added insight into why some quantum systems violate the Cerf-Adami inequalities thereby improving our understanding of the quantum-classical boundary.  As a result we suggest that the second law may serve as some type of boundary condition on classical knowledge.
\end{abstract}


\maketitle
\section{Introduction}
It has been argued that the laws governing entanglement may well be thermodynamic in nature, or, at the very least, possess thermodynamic corollaries \cite{horodeckioppenheim02, horodeckioppenheim03}.  For example, entanglement has been shown to be necessary in order for the third law of thermodynamics to be consistent with quantum theory \cite{brukner}.  At the heart of entanglement is the notion that the quantum states of two or more objects may be correlated in some way.  In 1964, Bell derived an upper bound on the classical strength of these correlations \cite{bell, bellspeak} and, since then, numerous experiments have proven that quantum correlations have strengths that exceed this upper bound \cite{aspect1, aspect2, kwiat95}.  BellÕs derivation and subsequent improvements on his original work have utilized correlation coefficients and expectation values as a measure of entanglement \cite{kwiat95, kwiat99}.  In 1996, Cerf and Adami introduced the use of entropy as a measure of entanglement and derived an upper bound on the strength of classical correlations using this measure \cite{cerfadami97a, cerfadami98}.  Entanglement plays a central role in quantum information theory \cite{nc} and entropy has long been a measure of information in classical information theory, having been formally introduced by Shannon \cite{shannon}, thus the step taken by Cerf and Adami was a natural one.

The importance of the Cerf-Adami inequalities, as we will call them, lies squarely in the fact that entropy \emph{is} a measure of information.  Note that some authors interpret information theoretic entropy in ways that are viewed as more consistent with the thermodynamic (Gibbs-Boltzmann) definition of entropy.  For example, Nielsen and Chuang refer to it as the amount of uncertainty that is present in a physical system.  However, they note that this makes it ideal for quantifying the resources required to store information \cite{nc}.  So, however we might look at it, entropy either quantifies the information storage capacity of a system or how much information we are able to \emph{access} about that system.  The Cerf-Adami inequalities utilize a certain type of entropy known as relative entropy that measures information about multiple systems or sub-systems at the same time.  For example, suppose we have a tripartite system in which there is a certain amount of information we might have if we knew parts \emph{A} and \emph{B} but not \emph{C}.  Conversely there is a certain amount of information we might have if we knew parts \emph{B} and \emph{C} but not \emph{A}, and likewise for \emph{A} and \emph{C} but not \emph{B}.  The Cerf-Adami inequalities essentially quantify the relationship of these relative entropies and thus compare the amount of information one might obtain about the system depending on which sub-systems one samples.

The discovery of the Cerf-Adami inequalities proved important for another reason.  They pointed to the need for a quantum analogue to the conditional entropy, i.e. a conditional von Neumann entropy.  Indeed, it was in [8] that this quantity was \emph{defined}.  In addition, they proved to be a generalization of the Braunstein-Caves inequality \cite{braunsteincaves} and thus have proven to be useful in understanding numerous information theoretic problems, e.g. quantum cryptographic protocols \cite{bechmann}.

Oddly enough, however, it turns out that the Cerf-Adami inequalities can be derived \emph{without} reference to conditional entropies.  The trouble with conditional entropies (and likewise conditional probabilities) stems from the fact that one \emph{could} interpret them as implying a time-like structure, e.g. \begin{math}H(B|A)\end{math}, the entropy of \emph{B} conditional on knowing \emph{A}, might be interpreted as implying some knowledge of \emph{A} must \emph{precede} this knowledge of \emph{B}.  Evidence for this interpretation appears in Cerf and Adami's original paper where they "define the conditional entropy \begin{math}H(A|B)\end{math} as the entropy of variable \emph{A} while "knowing", [sic] i.e., having measured, \emph{B}" \cite{cerfadami97a} which, as worded, implies a previous action.  This interpretation is strongly opposed by some \cite{terry} thus, a derivation of the Cerf-Adami inequalities that is free of conditional entropies also rids us of at least one debate.

There is also a similar debate concerning Markov chains.  While the latter are not \emph{explicitly} utilized in most derivations of the Cerf-Adami inequalities, as we will show they certainly are \emph{implicitly} utilized.  Thus, again, ridding ourselves of the need for Markov chains frees us from another interpretational point, further clearing up the meaning of these inequalities.  By then suggesting a potential use for these inequalities in experimentally probing the quantum-classical boundary, we ease the interpretational strain on any possible results.

In order demonstrate all of the nuances inherent in derivations of the Cerf-Adami inequalities, we walk through a derivation based on mutual information that does not include conditional entropies but does include Markov chains.  We then present an even simpler derivation that includes \emph{neither} conditional entropies \emph{nor} Markov chains.  We encourage the interested reader to compare these to Cerf's and Adami's original paper, in which there is no explicit mention of Markov chains.  The first derivation we give below demonstrates why Cerf's and Adami's original derivation \emph{implicitly} relies on Markov chains.

\section{The Cerf-Adami inequalities}
In information theory it is usual to represent entropy in the binary sense as articulated by Shannon \cite{shannon}, \begin{equation}H(X)=-\sum_{x}p(x)\,\textrm{log}\,p(x)\end{equation}
where the logarithm is taken to be base-two.\footnote{Note that the sign conventions employed throughout are consistent with those in \cite{nc}.}  Suppose we have two systems (or sub-systems) and we wish to measure over the indices \emph{x} and \emph{y}.  The \emph{joint} entropy measured over these indices is defined as
\begin{equation}H(X,Y)\equiv-\sum_{x,y}p(x,y)\,\textrm{log}\,p(x,y).\end{equation}  We define the relative entropy \cite{nc}, \begin{equation}\begin{split}H(p(x,y)\parallel p(x)p(y))&\equiv-\sum_{x,y}p(x,y)\,\textrm{log}\frac{p(x,y)}{p(x)p(y)}\\ & =H(p(x))+H(p(y))-H(p(x,y))\end{split}\end{equation} to be a measure of the "offset" of the probability distribution over two indices, \emph{x} and \emph{y}, from the probability distributions of the individual indices themselves.  As in \cite{nc}, we define \begin{math}-0\textrm{log}0\equiv0\end{math} and \begin{math}-p(x,y)\textrm{log}0\equiv+\infty\end{math} if \begin{math}p(x,y)>0\end{math}.  Since this represents an offset of the probability distributions, it is zero when these distributions are independent.

The relative entropy can be expressed in a number of ways including as the \emph{mutual} entropy that represents the mutual information of two systems.  As such, consider two systems, \emph{A} and \emph{B}, that are measured on indices \emph{x} and \emph{y} respectively.  For convenience (and for ease of transition later) we will dispense with the indices and simply refer to the entropy of the two systems as \begin{math}H(A)\end{math} and \begin{math}H(B)\end{math} respectively.  We can thus define the mutual entropy as \begin{equation}H(A:B)\equiv H(A)+H(B)-H(A,B).\end{equation}  See Figure 1 for a visual representation of this and note that \begin{math}H(A:B)\equiv H(A)\cap H(B)\end{math}.
\begin{figure}
\begin{center}
\includegraphics{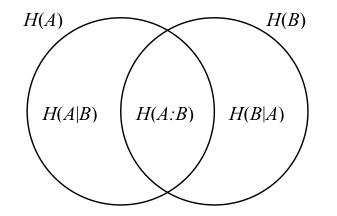}
\caption{In the language of set theory, \emph{H}(\emph{A}:\emph{B}) is the intersection of \emph{H}(\emph{A}) and \emph{H}(\emph{B}).  Also note that while it is standard practice to interpret \emph{H}(\emph{A}$\mid$\emph{B}) in such a way as to imply some sort of temporal order (see below), in purely set theoretic terms this is not necessary.}
\label{ }
\end{center}
\end{figure}
Note that equations (3) and (4) imply that \begin{equation}H(A:B)=H(p(x,y)\parallel p(x)p(y))\ge 0\end{equation} where the equality holds only when \emph{A} and \emph{B} are taken to be \emph{independent} systems measured over independent random variables \emph{x} and \emph{y} (that is \emph{p}(\emph{x,y})=0).

We define a Markov chain as an \emph{ordered} sequence, \begin{math}X_1 \rightarrow X_2 \rightarrow \cdots\end{math} of random variables such that \begin{math}X_{n+1}\end{math} is independent of \begin{math}X_{1}, \ldots, X_{n-1}\end{math} given \begin{math}X_n\end{math} \cite{nc}.  As such a Markov chain, as defined here, inherently contains the assumption that \begin{math}X_n\end{math} occurs \emph{before} \begin{math}X_{n+1}\end{math}.  Note that a frequent interpretation extends this such that a series of \emph{measurements} of such variables is also considered a Markov chain \cite{nc}.

Consider now \emph{three} systems, \emph{A}, \emph{B}, and \emph{C}, each having a corresponding entropy \begin{math}H(A)\end{math}, \begin{math}H(B)\end{math}, and \begin{math}H(C)\end{math}.  Suppose we measure the variable \begin{math}X\end{math} on these systems in such a way that the string of measurements on systems \begin{math}A\Rightarrow B\Rightarrow C\end{math} is a Markov chain.  Then it happens to also be true that the string of measurements on \begin{math}C\Rightarrow B\Rightarrow A\end{math} is also a Markov chain.  In essence there is a temporal order here that is assumed for the measurements of the variable on the systems where each measurement is considered to be independent of any of the others. \footnote{As it turns out, the very nature of Markov chains is somewhat debatable as evidenced by the numerous sometimes contradictory definitions one finds in the literature, e.g. \cite{penrose, landau, nc}.}

With these three systems we may also have \begin{math}H(B:C)\end{math} and \begin{math}H(A:C)\end{math} in addition to \begin{math}H(A:B)\end{math}.  As Nielsen and Chuang point out, the so called \emph{data processing inequality} supplies an information theoretic description of the conditions under which a Markov chain "loses" information about its earlier values as time progresses.  This inequality may be written (in terms of our systems, \emph{A}, \emph{B}, and \emph{C}),\begin{equation}
\begin{split}
H(A) \ge H(A:B) \ge H(A:C) \\
H(C) \ge H(C:B) \ge H(C:A)
\end{split}
\end{equation}
where the former is when the chain begins at \emph{A} and the latter is when the chain begins at \emph{C}.  In other words, if we begin with a certain amount of information about system \emph{A}, quantified as \begin{math}H(A)\end{math}, when we proceeded to systems \emph{B} and \emph{C} in order, we \emph{lost} information about system \emph{A}.  Essentially \begin{math}H(A)\end{math} gives us an upper bound on how much \emph{total} information we may possess.  Once we gain information about system \emph{B}, for instance, we must lose \emph{at least} an equivalent amount of information about \emph{A} such that the total mutual information we have can't exceed our predetermined limit.  It is clear from this that there is a time-like progression inherent in this description and that it implies a triangle inequality, \begin{equation}H(A:B)+H(B:C) \ge H(A:C)\end{equation} where this is an equality if no information exists about \emph{B}, i.e. \begin{math}H(B)=0\end{math}.

If these entropies of the individual systems are normalized (and we are still working with the Markov chain assumption), then the mutual information is symmetric, i.e. \begin{math}H(A:B)=H(B:A)\end{math}, something that should be evident from Figure 1.  Equations (6) and (7) together, then, imply 
\begin{equation}
\begin{split}
H(A:B)+H(B:C)-H(A:C) \le H(B) \\
H(A:B)+H(A:C)-H(B:C) \le H(A) \\
-H(A:B)+H(A:C)+H(B:C) \le H(C).
\end{split}
\end{equation}
Further, if the indices that the systems are measured on represent uniform distributions, then \begin{math}H(A)=H(B)=H(C)=1\end{math} and we arrive at the Cerf-Adami inequalities, \begin{equation}|H(A:B)-H(A:C)|+H(B:C) \le 1\end{equation} that have been shown to be in perfect analogy to the usual form of Bell's inequalities and that greatly resemble the Braunstein-Caves inequalities \cite{cerfadami97a, braunsteincaves}.

Why, again, are these inequalities important?  First, they serve as a way to define properties of the conditional entropy \cite{cerfadami97a, cerfadami98} and put bounds on the sharing of information between systems (parties) thus proving useful for analysis of quantum cryptographic protocols \cite{bechmann}.  In addition, violations of these inequalities by certain quantum systems is an indication of the non-separability of those quantum systems \cite{cerfadami97b}.  Due to the ubiquity of entropy as a measure in information theory, it makes sense that a set of Bell inequalities based on entropy would be more useful than the original set which are based on correlation coefficients and expectation values.

In any case, we have succeeded here in demonstrating a derivation of the Cerf-Adami inequalities that utilizes Markov chains but not conditional entropies, even though the latter were defined by the original paper that introduced these inequalities.  We have also discussed the (potential) temporal nature of Markov chains and conditional entropies.  Note that the Shannon entropy is positive definite.  Penrose has rigorously shown that any non-decreasing statistical entropy must satisfy two additional constraints: the Markov chain must be \emph{deterministic}, and only the \emph{number} of individual systems can be observed and not their identities \cite{penrose}.  It seems as if the temporal order is a clear requirement, particularly if Markov chains are employed.  But what if Markov chains were \emph{not} employed?  We shall now show an alternative derivation that does not involve Markov chains and, in fact, makes no reference to any temporal order, either explicitly or implicitly.  In fact it avoids an interpretation of the conditional entropy altogether by not using it.  Perhaps oddly, we turn to statistical mechanics for this, though the latter is often associated with a temporal order.

\section{An alternative treatment}
There is an alternative definition for the entropy that is commonly used in statistical mechanic and thermodynamic situations that we now introduce.  In order to fully articulate it we must first introduce the concept of multiplicity.  In most isolated systems there are usually many ways in which the system may configure itself in order to achieve a single macroscopic state, sometimes called a \emph{macrostate}.  Each of the ways in which the system may configure itself is usually called a \emph{microstate}.  So each macrostate usually consists of several microstates.  For example, consider a pair of dice.  A roll of 7 would constitute a macrostate.  There are six ways in which we might achieve a roll of 7 (assuming the dice are classical) thus there are six microstates associated with the given macrostate.  The so-called fundamental assumption of statistical mechanics assumes that all six of these microstates for the roll of 7 are equally probable in the long run.  In thermodynamic systems such as a gas an ensemble that has many microstates is often called a microcanonical ensemble.  The multiplicity, \begin{math}\Omega\end{math}, of a state is simply the number of microstates for that given macrostate (e.g. it would be six for a roll of 7 on a pair of dice).

In real thermodynamic systems it is often the case that the multiplicity is an enormous number (e.g \begin{math}10^{120}\end{math}.  Thus, it is often easier and more desirable to logarithmically scale this value.  In fact we often define the entropy as \begin{equation}H=\frac{S}{k_\textrm{B}}=\textrm{ln}\,\Omega\end{equation} where \begin{math}k_\textrm{B}\end{math} is Boltzmann's constant.\footnote{Note that thermodynamicists usually denote entropy as \emph{S} rather than \emph{H}.  Information theorists often attribute \emph{S} to the von Neumann entropy while \emph{H} is the Shannon entropy.  Since the latter is classical we will denote any classical entropy with \emph{H} for consistency.}  This is usually called the Boltxmann entropy and is entirely equivalent to equation (1) since the base in both cases is arbitrary (see \cite{schroeder} for a proof of this equivalence\footnote{Information theorists often use the Shannon entropy and thus base-two.  Thermodynamicists tend to use base-ten.  In truth, since the logarithm merely scales the multiplicity, it is somewhat arbitrary.  The units of entropy in (12) are introduced via BoltzmannÕs constant and it is not uncommon to actually use the unitless value $S/k_\textrm{B}$ instead.  Thus, the Shannon entropy, which is technically unitless (though often assigned units of ÔbitsÕ by information theorists) can easily be converted to Boltzmann entropy and vice-versa if one converts the multiplicities to probabilities (again, see \cite{schroeder}).  Recall that the standard logarithm conversion is $\textrm{log}_a(x)=\textrm{log}_b(x)/\textrm{log}_b(a)$.}).  It is quite clear, given the definition of multiplicity, that this definition of entropy is positive definite.

\subsection{Combinations of systems}
Now consider two systems, \emph{A} and \emph{B}, with multiplicities \begin{math}\Omega(A)\end{math} and \begin{math}\Omega(B)\end{math}.  Since multiplicity counts microstates, if these systems are combined we would expect the combined systemÕs multiplicity to be a \emph{product} of \begin{math}\Omega(A)\end{math} and \begin{math}\Omega(B)\end{math}.  For example, say system \emph{A} is a pair of dice showing 7 and system \emph{B} is a pair of dice showing 8.  The multiplicities are six and five respectively.  Thus the multiplicity of a roll of 15 on \emph{four} dice is thirty Ð the multiplicities are multiplicative.  The behavior of thermodynamic systems is generally consistent with this idea \cite{penrose, schroeder, landau}.  So, for example, for systems \emph{A} and \emph{B} once they are combined (or considered in unison), the multiplicity of the combination would be
\begin{equation}
\Omega(A,B)=\Omega(A)\cdot\Omega(B)=e^{H(A)}\cdot e^{H(B)}=e^{H(A)+H(B)}.
\end{equation}
The total entropy, then, is seen to be additive since
\begin{equation}
H(A,B)=H(A)+H(B)=\textrm{ln}\,\Omega(A,B).
\end{equation}
Now this is a very simplified example.  In real thermodynamic systems the multiplicity is often a complicated function of volume, number of molecules, temperature, pressure and other thermodynamic quantities.  Nonetheless, the additivity of classical entropy is well-established \cite{penrose, schroeder, landau} and, in fact, entropy is actually \emph{sub}additive \cite{nc}, meaning 
\begin{equation}
H(A,B)\le H(A)+H(B)
\end{equation}
since, in some cases, one can imagine certain microstates might be redundant or might combine (in fact the subadditivity of the entropy is what leads to equation (4) where the equality in (15) holds if they are independent and the mutual entropy is zero).

\subsection{Counting bits}
Note that Shannon entropy, while technically unitless, is often ÔmeasuredÕ in bits, i.e. the number of bits of information for a given system.  Since we have divided by BoltzmannÕs constant in (12) in order to make it unitless, there is nothing preventing us from doing the same with the Boltzmann entropy.  So suppose we have two systems, \emph{A} and \emph{B}, about which we have some information in the form of entropy counted in bits.  Suppose further that some of these bits of information actually tell us something about \emph{both} systems simultaneously so these bits count as entropy for both systems.  These bits are analogous to a person with dual citizenship who is counted in both his or her countries' censuses.  \begin{math}H(A)\end{math}, then, counts \emph{all} the bits that tell us something about system \emph{A}.  Likewise, \begin{math}H(B)\end{math} counts all the bits that tell us something about system \emph{B}.  The bits that gives us information about \emph{both} systems are technically counted twice, then, since they are included in both \begin{math}H(A)\end{math} and \begin{math}H(B)\end{math}.  By themselves, these bits are labeled \begin{math}H(A:B)\end{math} since they represent information about \emph{both} systems.  The total number of bits we have, that is the joint entropy \begin{math}H(A,B)\end{math}, is
\begin{equation}
H(A,B)=H(A)+H(B)-H(A:B)
\end{equation}
where we subtract off \begin{math}H(A:B)\end{math} once so the bits with information about both systems don't get counted twice.

Now consider a third system \emph{C}.  It is trivially true that
\begin{equation}
H(A,B)+H(B,C)\ge H(A,C)
\end{equation}
where the equality holds if we have no information about system \emph{B}, i.e. \begin{math}H(B)=0\end{math}.  When we combine this with equation (14), which is simply a rearrangement of equation (4), we find that
\begin{equation}
\begin{split}
&\{H(A)+H(B)-H(A:B)\}\\&+\{H(B)+H(C)-H(B:C)\} \\ &\ge H(A)+H(C)-H(A:C).
\end{split}
\end{equation}
Reducing and rearranging this produces
\begin{equation}H(A:B)+H(B:C)-H(A:C) \le 2H(B).
\end{equation}
It turns out that we may further narrow this bound.  Suppose we have a total of \emph{n} bits equally distributed among our three systems such that \begin{math}H(A)=H(B)=H(C)=n/3\end{math} and \begin{math}H(A)+H(B)+H(C)=n\end{math}.  Let us assume that it is \emph{not} possible for, say, bits from system \emph{A} to give information about system \emph{B} but not the reverse.  This means \begin{math}H(A:B)=H(B:A)\end{math}.  Given that, suppose \emph{all} of the bits in \emph{A} also give us information about \emph{B}.  Our previously stated condition requires the reverse to be true.  In this case, the total number of bits with "dual citizenship" is \begin{math}H(A:B)_{\textrm{max}}=H(A)+H(B)=2n/3\end{math}, or, in the non-maximal case, \begin{math}H(A:B) \le 2n/3\end{math}.  Suppose the same is true for systems \emph{B} and \emph{C}.  If that were the case, \begin{math}H(B:C) \le 2n/3\end{math}.  Suppose \emph{one} of these two is at a maximum, e.g. \begin{math}H(A:B)_{\textrm{max}}=2n/3\end{math}.  Since we only have a total of \emph{n} bits to work with, this limits \begin{math}H(B:C)\end{math} to a maximum of \emph{n}/3.  Adding a third group of shared bits, \begin{math}H(A:C)\end{math}, further reduces this limit.  However, by introducing this third group of shared bits we have introduced the possibility of having bits that give us information about \emph{all three} systems.  Bits of this sort may be labeled \begin{math}H(A:B:C)\end{math}, but note that we run the risk of counting these bits \emph{three times} since they appear in \begin{math}H(A)\end{math}, \begin{math}H(B)\end{math}, and \begin{math}H(C)\end{math}.  Thus \begin{equation}H(A)=H(B)=H(C) \le n/3\end{equation} and \begin{equation}H(A)+H(B)+H(C)-2H(A:B:C)=n.\end{equation}

Suppose \begin{math}H(A:B)\end{math} is at a maximum.  That means that there is no way to distinguish between the bits of system \emph{A} and those of system \emph{B} and thus \begin{math}H(A:B)_{\textrm{max}}=H(A)=H(B)\end{math}.  This further implies that \begin{math}H(A:B:C)\end{math}, \begin{math}H(A:C)\end{math}, and \begin{math}H(B:C)\end{math} \emph{all represent the exact same set of bits} meaning their labels are interchangeable.  In other words, in this case, \begin{equation}\begin{split}H(A:B)_{\textrm{max}} &\Rightarrow\\ &H(A:B:C) \equiv H(A:C)\\ &\equiv H(B:C)\end{split}\end{equation} where we read \begin{math} \equiv\end{math} as "is identical to" rather than "is defined by" or "is equivalent to" since it means they truly are the same set of bits.  Note that it is also true that \begin{math}H(A) \equiv H(B) = 2n/3\end{math}. In any case, these arguments first imply that \begin{math}H(A:B)_{\textrm{max}} \le 2H(B)\end{math} and then, because \begin{math}H(A:B)_{\textrm{max}}=H(B)\end{math}, they further imply we may drop the factor of 2 as being redundant.  This works regardless of which systems we maximally combine since the letters are merely labels for sets of bits.  As such we may further narrow the bound on equation (17) and write
\begin{equation}
H(A:B)+H(B:C)-H(A:C) \le H(B).
\end{equation}
Furthermore, when \begin{math}H(A)\end{math},  \begin{math}H(B)\end{math}, and  \begin{math}H(C)\end{math} represent uniform probability distributions, their normalization can be set to unity.  Likewise, we could permute the letters depending upon which systems we are comparing.  Thus thus may be generalized to equation (9) which are the Cerf-Adami inequalities,
\begin{equation}
|H(A:B)-H(A:C)|+H(B:C) \le 1. \tag{9}
\end{equation}

\section{Entropy}
As Landau and Lifshitz point out \cite{landau}, there are inherent difficulties in the interpretation of entropy in terms of the units (i.e. the units are either entirely wrapped up in the multiplicative constant, \begin{math}k_{\textrm{B}}\end{math} or the ÔunitsÕ of bits are assigned to what is technically a unitless quantity such as the Shannon entropy).  In fact there are numerous problems inherent in the concept of entropy (see, for example, [18, 19]).  As such, the only \emph{uniquely} determined quantities that do not depend on the choice of units are \emph{differences} in entropy, i.e. the changes in entropy brought about by some process \cite{landau}.  Consider, for example, two systems that are initially separated and then allowed to interact in some manner (for example, two ideal gases separated by a barrier that is later removed).  For classical systems, the total entropy of the combined system after mixing is always the same as or greater than it was \emph{before} mixing \cite{schroeder}.  In other words, this change in total entropy, often called the \emph{entropy of mixing} \cite{schroeder}, is always positive, i.e.,\begin{equation}
H_{\textrm{mix}}=\Delta H_{\textrm{total}} \ge 0
\end{equation}
where the equality holds if the systems are identical (e.g the same type of gas).  For example, then, if the systems represent ideal gases and entropy is a method for expressing the probability that a system will be in a given state, the individual entropies provide a method for expressing the probability distributions of the two systems.  If the two gases were the same species and otherwise identical prior to mixing, there would be no difference between the two probability distributions and thus no entropy of mixing (technically the multiplicity increases slightly but the factor is negligible and thus it is approximately zero, but always positive regardless).

How might we explain this in terms of bits?  Consider two systems of bits, \emph{A} and \emph{B}.  Note that if there is no mixing, i.e. no bits with 'dual citizenship' (mutual information), then \begin{math}H(A:B)=0\end{math}.  As the number of bits with information about both systems increases, \begin{math}H(A:B)\end{math} increases.  Notice also that this quantity can never be negative \emph{even if we try pulling the systems apart}.  It \emph{can decrease}, but it can never be less than zero.  Thus the mutual entropy is very similar to the entropy of mixing.  It may seem the analogy isnÕt quite perfect since, in the thermodynamic case the change is in the total entropy while in the information theoretic case the total number of bits appears to remain the same.  But there is nothing in the information theoretic case preventing the 'creation' of bits by some other process like noise, for instance.  So in thermodynamic systems such as the example of mixing two ideal gases, while \begin{math}H(A:B)\end{math} increases, we might expect \begin{math}H(A)\end{math} and \begin{math}H(B)\end{math} to correspondingly increase which, in fact they do since entropy is a function of volume and by removing the barrier the gases now each have a greater volume through which to spread.  Thus we interpret the mutual entropy as a generalization of the entropy of mixing.  The mixing process for three systems is described by (14).  The entropy of mixing is sometimes interpreted as the work required to mix the systems, but the notion of work is as fraught with problems [20] as that of entropy (perhaps moreso since the problems are largely taxonomic).  Either way, we see that changes in the total entropy for isolated systems is never negative (i.e. it never decreases).  When pressed for a mathematical statement of the second law of thermodynamics, this is the answer that is frequently given, though usually in a form similar to \begin{math}\delta S_{\textrm{isolated}} \ge 0\end{math} where most thermodynamicists use \emph{S} for entropy.  In essence, then, the positivity of the mutual information, argued heuristically a moment ago and clarified in equation (5), is a statement of the second law of thermodynamics.  Thus, it is quite clear the the Cerf-Adami inequalities are intimately related to and perhaps even dependent upon the second law of thermodynamics.

As a brief historical note, since the late 1950s, no less than fifteen articles proposing new or revised statements of the second law have appeared \emph{in a single journal}, that being the \emph{American Journal of Physics}.  These include a generalized form of the second law of thermodynamics in terms of information that appeared in 1964 \cite{rodd} and a form \emph{derived from quantum mechanics} that appeared in 1965 \cite{wannier}!  The most recent "new" statement of this law in \emph{Am. J. Phys.} appeared in 1995 \cite{macdonald} while, in 1997, Moore and Schroeder argued in favor of a version (not necessarily new) based on probabilities and multiplicities that is similar to the simple argument we give below examples using coins \cite{mooreschroeder}.  Other traditional statements include the well-known version of Kelvin and Planck, that of Clausius, and another known as the Sears-Kestin statement of the second law (see, for example, \cite{lu}).  Simply put, to this day agreement on a statement of the second law, particularly a mathematically quantifiable one, is strongly debated.  As Partovi recently pointed out, "rarely have so many distinguished physicists written as extensively on a subject while achieving so little consensus" \cite{partovi}.

\section{Consequences}
Thusfar we have demonstrated a relationship between the Cerf-Adami inequalities and the second law of thermodynamics.  We have also demonstrated the the aforementioned inequalities can be derived without reference to either Markov chains or conditional entropies.  There are several points of significance to this.

\subsection{Conditional entropies and entanglement}
Bell-type inequalities are often used experimentally to measure entangled states.  In fact certain forms of entropy can be used as a measure of (i.e. to quantify) entanglement.  To see this, let us first define the von Neumann as \begin{equation}S(\rho) \equiv -\textrm{tr}(\rho \, \textrm{log} \, \rho)\end{equation} where \begin{math}\rho\end{math} is a density operator that represents the quantum state of the system.  Note that we use \emph{S} by convention to differentiate it from the classical entropy, \emph{H}.  

The joint entropy of a system with two components, \emph{A} and \emph{B}, is \begin{equation}S(A,B)\equiv -\textrm{tr}(\rho^{AB}\,\textrm{log}\,(\rho^{AB})).\end{equation}  The \emph{quantum} conditional entropy, that is the quantum entropy of \emph{B} conditional upon knowing \emph{A}, is then defined as \begin{equation}S(B|A)\equiv S(A,B) - S(A).\end{equation}  If \begin{math}|AB\rangle\end{math} is a pure state of a composite system, \begin{math}|AB\rangle\end{math} is entangled \emph{if and only if} \begin{math}S(B|A) < 0.\end{math}  To quote directly from Cerf's and Adami's original paper, "a violation of an entropic Bell inequality always goes hand in hand with the appearance of a negative conditional entropy" at least from the Venn diagram perspective (\cite{cerfadami97a}, p.3).  But that is precisely the perspective from which we have derived these inequalities \emph{without} any reference to conditional entropies.  Clearly, then, the violation of these inequalities must have its roots elsewhere.  Since conditional entropies essentially rescale probability distributions, this seems to imply that violations of Bell inequalities are not necessarily related to conditional probability distributions.

\subsection{Temporal evolution}
Deriving the Cerf-Adami inequalities using Markov chains seems to imply there is some temporal order inherent in the inequalities themselves, particularly when considering equation (7) that proceeds directly from equation (6) and the Markov chain assumption.  One might be tempted to assume that conditional entropies \emph{also} imply some sort of temporal order, though this is not a universally accepted interpretation of said entropies.  Either way it doesn't matter since we have demonstrated a derivation of these inequalities without reference to \emph{either}.  This does not necessarily mean that there is no temporal order or evolution inherent in these inequalities.  It simply means any such order would have to be associated with something other than the Markov chains and/or conditional entropies.  If there \emph{was} such an order inherent in these inequalities, where might it be?

Consider that we have argued that the mutual entropies are a representative statement of the second law of thermodynamics.  The second law has often been associated with the arrow of time (though, Partovi has recently demonstrated a reversal of this arrow in macroscopic systems is possible \cite{partovi}).  Thus it might be possible to trace a temporal order to the \emph{mutual} entropy (information).  In other words, the temporal order arises from the positivity of the mutual entropy, or, more colloquially, once two systems are mixed it's nearly impossible to perfectly separate them.  In fact, one can, of course, speak directly in terms of probability distributions here since that is really what entropies measure.  In the Cerf-Adami inequalities, this implies that there is a specific order taken when accessing or measuring the systems.

What does a violation of these inequalities mean then in terms of the second law and the arrow of time?  Does it mean that violations of these inequalities implies some violation of causality?  Actually, it doesn't and here is why.  

\subsubsection{Reversibility}
Ultimately the second law is tied to the idea of reversibility.  Consider the following two simple examples.\\\\
\noindent\textbf{Example 1} \emph{Suppose we have a single coin.  The probability of tossing it and having it land with its head} (H) \emph{showing, is 0.5.  The same is, of course, true for its tail} (T).  \emph{Suppose we toss it twice and we get} H-T.  \emph{So now it is laying with it's tail showing on the back of our hand.  Suppose we want to reverse this process, i.e., since it is presently in state} T, \emph{we wish to get it back to state} H.  \emph{The probability of doing so is, of course, 0.5.  But now, suppose we toss it five times in a row and we get} H-T-H-H-T.  \emph{Say we wish to reverse} this \emph{process, that is we want} T-H-H-T-H.  \emph{The probability of accomplishing} that \emph{is only 0.03125!}\\\\
\noindent\textbf{Example 2} \emph{Suppose now we have five coins we wish to flip} simultaneously. \emph{Say we do so and the result is} H-T-H-H-T.  \emph{Say we do so again and the result} T-H-T-T-H.  \emph{Suppose we want to reverse this process perfectly, that is, starting with} T-H-T-T-H \emph{showing, we wish to flip the coins such that they return to the state} H-T-H-H-T.  \emph{The probability of accomplishing that is} also \emph{only 0.3125!}\\\\
\noindent  What these two examples show is that a) the further we have progressed through a series of singular random processes, the harder (less probable) it becomes to reverse the series exactly and b) the larger a system is, the harder (less probable) it becomes to reverse a \emph{single} process.  This is the heart of the second law of thermodynamics.  Many microscopic processes are perfectly reversible, but no \emph{macroscopic} process is (or so we thought until Partovi's recent work \cite{partovi}).  Macroscopic processes may be \emph{approximately} reversible (e.g. the action of opening and closing a door), but it is important to remember that this is only an approximation (open and close the door enough and you'll introduce wear).  In this way we might argue that the second law is a strong statement concerning the nature of probabilities and aggregate systems.  Indeed, Schroeder has argued this very point \cite{schroeder}.  So one way to view the second law and the origin of the arrow of time is as a consequence of constructing macroscopic systems out of many microscopic systems.  It's essentially related to the law of large numbers.  In our derivations above, we always assumed our distributions were normalized, and thus you'd expect that if the entropies obey the inequality, the probabilities do and the relation of the probabilities to \emph{each other} should not depend on the size of the system.  One might then argue that the arrow of time is a result of the inherent tendency of the constituents of the universe to "clump" to form macroscopic objects (we use macroscopic in a broad sense that includes complex molecules, for example).  Of course, the discovery of macroscopically entangled and perfectly reversible systems by Partovi adds another element to this argument that we discuss below.

\subsubsection{An epistemic interpretation}
There is another way to look at this, however.  We can view relative entropies as providing us with epistemic information about the systems that are involved in the problem.  So for instance, \begin{math}H(A:B)\end{math}, the mutual information of systems \emph{A} and \emph{B}, could be viewed as a mutual boundary condition on both \emph{A} and \emph{B}.  So equation (7), for instance, establishes an ordered set of boundary conditions on the knowledge we have (or may obtain) concerning these systems.  The second law is sometimes interpreted as being the fact that the universe has an initial boundary condition but not necessarily a final boundary condition.  Thus the Cerf-Adami inequalities may be interpreted as placing boundary conditions on the \emph{classical} knowledge we may obtain about a system.  In this sense it represents the limit to which our classical knowledge may take us and, beyond which, lies the quantum world whose information is a bit different.  Since knowledge may be measured as information in the form of entropies, the generality of the von Neumann entropy arises naturally from this description, i.e. while the von Neumann entropy can \emph{reduce} to a classical entropy in certain situations, it is more general in that it allows for negative probabilities (or, rather, non-seperable density operators).  In other words the von Neumann entropy \emph{extends} our knowledge beyond its usual limit.  Now, technically this has nothing to do with the second law nor with the thermodynamic arrow of time.  As such, it does not necessarily seem immediately true that a violation of the Cerf-Adami inequalities implies the existence of a non-causal process.  However, it may be that the second law itself and, by extension the thermodynamic arrow of time, are both simply \emph{boundary conditions on classical knowledge}.  In other words, classical information is completely causal while it might be possible that \emph{some} quantum information is not or at least does not have as strict a set of boundary conditions.  In other words, classical information, it seems, is entirely governed by initial boundary conditions whereas quantum information seems to be a bit looser in that it could be governed, at least partially, by some unknown \emph{final} boundary condition.

\subsection{The quantum-classical boundary}
So, is the microscopic-macroscopic "boundary" discussed in the previous subsection the same as the quantum-classical boundary?  Certainly we are not accustomed to thinking of \emph{macroscopic} quantum processes and thus, perhaps, we are inclined to equate the two boundaries.  But it may not necessarily be true that they are one and the same.  It was Schr\"{o}dinger's contention that the signature of a quantum system was one that is non-seperable (non-factorable), i.e. entangled (see \cite{loepp}).  Entanglement, as it turns out, has absolutely nothing to do with the size of the system and macroscopic entanglement was recently suggested by Partovi as being associated with so-called ambient correlations \cite{partovi}.  Nor does it necessarily have anything to do with any lengthy string of singular processes such as the first example we gave in the previous section.  Understanding the quantum-classical boundary, then, means developing a better grasp of entanglement and \emph{not} necessarily comparing microscopic and macroscopic systems.  Would the Cerf-Adami inequalities, then, provide us with a way of probing this boundary?  Generally, one might be inclined to think so.  But note that there is nothing that\emph{requires} quantum systems violate these inequalities.  If we have unentangled states we don't necessarily expect these inequalities to be violated.

What's happening here?  Well, the confusion comes from the fact that we refer to quantities such as the von Neumann entropy as being "quantum" when, in reality, the von Neumann entropy is really just a generalized entropy that could be used for \emph{any} system, even a macroscopic, classical one, since, in such a case, it just reduces to the classical entropy.  In addition, the language that we refer to as quantum mechanics, might better be thought of as \emph{microscopic} mechanics if we are to take Schr\"{o}dinger's view that quantum systems are entangled systems.  In a sense, it appears there might be a slight semantic difference that leads to larger conceptual problems.  In other words, the Cerf-Adami inequalities do not necessarily get us that much closer to understanding precisely what it is that makes quantum states and systems unique.

Note that, while it might be tempting to consider the uncertainty relations as another sign of "quantumness," the generalized form of those relations as developed by Schr\"{o}dinger and Robertson are actually purely mathematical relations between certain types of operators and are completely independent of anything "quantum."

\begin{acknowledgements}
We are extremely indebted to Terry Rudolph, Ken Wharton, Barry Sanders, and Frank Schroeck for assistance with several versions of this manuscript.  The immense amount of time they each devoted to scrutinizing this work was tremendously helpful.  Additionally we thank Paul OÕHara, Caslav Brukner, Greg Buck, and Ali Rezakhani for discussion, support, insight, and helpful comments.
\end{acknowledgements}

\bibliographystyle{apsrev}
\bibliography{CerfAdamiIneq.bbl}

\end{document}